\documentstyle[12pt]{article}
\newcommand{\nc}{\newcommand}
\nc{\renc}{\renewcommand}

\nc{\half}{{\textstyle{1\over2}}}
\nc{\etal}{\mbox{\it et al.}}
\nc{\ie}{{\it i.e.}}
\nc{\eg}{{\it e.g.}}

\renc{\thefootnote}{\arabic{footnote}}
\nc{\capt}[1]{{\bf Figure.} {\small\sl #1}}

\nc{\eqs}[2]{\mbox{Eqs.~(\ref{#1},\,\ref{#2})}}
\nc{\eq}[1]{\mbox{Eq.~(\ref{#1})}}

\nc{\figs}[2]{\mbox{Figs.~(\ref{#1},\,\ref{#2})}}
\nc{\fig}[1]{\mbox{Fig~.(\ref{#1})}}

\nc{\tag}[1]{\label{#1} \marginpar{{\footnotesize #1}}}
\nc{\mtag}[1]{\label{#1} \mbox{\marginpar{{\footnotesize #1}}}}
\renc{\baselinestretch}{1.2}
\newlength{\overeqskip}
\newlength{\undereqskip}
\nc{\be}[1]{\begin{equation} \mbox{$\label{#1}$}}
\nc{\bea}[1]{\begin{eqnarray} \mbox{$\label{#1}$}}
\nc{\Section}[2]{\section{#2}\label{#1}}
\nc{\Bibitem}[1]{\bibitem{#1}}
\nc{\Label}[1]{\label{#1}}

\nc{\eea}{\vspace{\undereqskip}\end{eqnarray}}
\nc{\ee}{\vspace{\undereqskip}\end{equation}}
\nc{\bdm}{\begin{displaymath}}
\nc{\edm}{\end{displaymath}}
\nc{\dpsty}{\displaystyle}
\nc{\bc}{\begin{center}}
\nc{\ec}{\end{center}}
\nc{\ba}{\begin{array}}
\nc{\ea}{\end{array}}
\nc{\bab}{\begin{abstract}}
\nc{\eab}{\end{abstract}}
\nc{\btab}{\begin{tabular}}
\nc{\etab}{\end{tabular}}
\nc{\bit}{\begin{itemize}}
\nc{\eit}{\end{itemize}}
\nc{\ben}{\begin{enumerate}}
\nc{\een}{\end{enumerate}}
\nc{\bfig}{\begin{figure}}
\nc{\efig}{\end{figure}}
\nc{\arreq}{&\!=\!&}
\nc{\arrmi}{&\!-\!&}
\nc{\arrpl}{&\!+\!&}
\nc{\arrap}{&\!\!\!\approx\!\!\!&}
\nc{\non}{\nonumber\\*}
\nc{\align}{\!\!\!\!\!\!\!\!&&}
\def\lsim{\; \raise0.3ex\hbox{$<$\kern-0.75em
      \raise-1.1ex\hbox{$\sim$}}\; }
\def\gsim{\; \raise0.3ex\hbox{$>$\kern-0.75em
      \raise-1.1ex\hbox{$\sim$}}\; }
\nc{\DOT}{\hspace{-0.08in}{\bf .}\hspace{0.1in}}
\nc{\Laada}{\hbox {$\sqcap$ \kern -1em $\sqcup$}}
\nc\loota{{\scriptstyle\sqcap\kern-0.55em\hbox{$\scriptstyle\sqcup$}}}
\nc\Loota{{\sqcap\kern-0.65em\hbox{$\sqcup$}}}
\nc\laada{\Loota}
\nc{\qed}{\hskip 3em \hbox{\BOX} \vskip 2ex}

\nc{\real}{{\rm I \! R}}
\nc{\Z}{{\sf Z \!\!\! Z}}
\nc{\complex}{{\rm C\!\!\! {\sf I}\,\,}}
\def\bigid{\leavevmode\hbox{\small1\kern-3.8pt\normalsize1}}
\def\id{\leavevmode\hbox{\small1\kern-3.3pt\normalsize1}}
\nc{\slask}{\!\!\!/}
\nc{\bis}{{\prime\prime}}
\nc{\pa}{\partial}
\nc{\na}{\nabla}
\nc{\ra}{\rangle}
\nc{\la}{\langle}
\nc{\goto}{\rightarrow}
\nc{\swap}{\leftrightarrow}

\nc{\EE}[1]{ \mbox{$\cdot10^{#1}$} }
\nc{\abs}[1]{\left|#1\right|}
\nc{\at}[2]{\left.#1\right|_{#2}}
\nc{\norm}[1]{\|#1\|}
\nc{\abscut}[2]{\Abs{#1}_{\scriptscriptstyle#2}}
\nc{\vek}[1]{{\rm\bf #1}}
\nc{\integral}[2]{\int\limits_{#1}^{#2}}
\nc{\inv}[1]{\frac{1}{#1}}
\nc{\dd}[2]{{{\partial #1}\over{\partial #2}}}
\nc{\ddd}[2]{{{{\partial}^2 #1}\over{\partial {#2}^2}}}
\nc{\dddd}[3]{{{{\partial}^2 #1}\over
        {\partial #2 \partial #3}}}
\nc{\dder}[2]{{{d #1}\over{d #2}}}
\nc{\ddder}[2]{{{d^2 #1}\over{d {#2}^2}}}
\nc{\dddder}[3]{{d^2 #1}\over
        {d #2 d #3}}
\nc{\dx}[1]{d\,^{#1}x}
\nc{\dy}[1]{d\,^{#1}y}
\nc{\dz}[1]{d\,^{#1}z}
\nc{\dl}[1]{\frac{d\,^{#1}l}{(2\pi)^{#1}}}
\nc{\dk}[1]{\frac{d\,^{#1}k}{(2\pi)^{#1}}}
\nc{\dq}[1]{\frac{d\,^{#1}q}{(2\pi)^{#1}}}

\nc{\cc}{\mbox{$c.c.$ }}
\nc{\hc}{\mbox{$h.c.$ }}
\nc{\cf}{cf.\ }
\nc{\erfc}{{\rm erfc}}
\nc{\Tr}{{\rm Tr\,}}
\nc{\tr}{{\rm tr\,}}
\nc{\pol}{{\rm pol}}
\nc{\sign}{{\rm sign}}
\nc{\bfT}{{\bf T }}

\def\GeV{{\rm\ GeV}}

\nc{\cA}{{\cal A}}
\nc{\cB}{{\cal B}}
\nc{\cD}{{\cal D}}
\nc{\cE}{{\cal E}}
\nc{\cG}{{\cal G}}
\nc{\cH}{{\cal H}}
\nc{\cL}{{\cal L}}
\nc{\cO}{{\cal O}}
\nc{\cT}{{\cal T}}
\nc{\cN}{{\cal N}}
\nc{\rvac}[1]{|{\cal O}#1\rangle}
\nc{\lvac}[1]{\langle{\cal O}#1|}
\nc{\rvacb}[1]{|{\cal O}_\beta #1\rangle}
\nc{\lvacb}[1]{\langle{\cal O}_\beta #1 |}
\nc{\bb}{\bar{\beta}}
\nc{\bt}{\tilde{\beta}}
\nc{\ctH}{\tilde{\cal H}}
\nc{\chH}{\hat{\cal H}}
\nc{\1}{\aa}
\nc{\2}{\"{a}}
\nc{\3}{\"{o}}
\nc{\4}{\AA}
\nc{\5}{\"{A}}
\nc{\6}{\"{O}}
\nc{\al}{\alpha}
\nc{\g}{\gamma}
\nc{\Del}{\Delta}
\nc{\e}{\epsilon}
\nc{\eps}{\epsilon}
\nc{\lam}{\lambda}
\nc{\om}{\omega}
\nc{\Om}{\Omega}
\nc{\ve}{\varepsilon}
\nc{\mn}{{\mu\nu}}
\nc{\ka}{\kappa}
\nc{\vp}{\varphi}

\nc{\advp}[3]{{\it  Adv.\ in\ Phys.\ }{{\bf #1} {(#2)} {#3}}}
\nc{\annp}[3]{{\it  Ann.\ Phys.\ (N.Y.)\ }{{\bf #1} {(#2)} {#3}}}
\nc{\apl}[3]{{\it  Appl. Phys. Lett. }{{\bf #1} {(#2)} {#3}}}
\nc{\apj}[3]{{\it  Ap.\ J.\ }{{\bf #1} {(#2)} {#3}}}
\nc{\apjl}[3]{{\it  Ap.\ J.\ Lett.\ }{{\bf #1} {(#2)} {#3}}}
\nc{\app}[3]{{\it Astropart.\ Phys.\ }{{\bf #1} {(#2)} {#3}}}
\nc{\cmp}[3]{{\it  Comm.\ Math.\ Phys.\ }{{ \bf #1} {(#2)} {#3}}}
\nc{\cqg}[3]{{\it  Class.\ Quant.\ Grav.\ }{{\bf #1} {(#2)} {#3}}}
\nc{\epl}[3]{{\it  Europhys.\ Lett.\ }{{\bf #1} {(#2)} {#3}}}
\nc{\ijmp}[3]{{\it Int.\ J.\ Mod.\ Phys.\ }{{\bf #1} {(#2)} {#3}}}
\nc{\ijtp}[3]{{\it Int.\ J.\ Theor.\ Phys.\ }{{\bf #1} {(#2)} {#3}}}
\nc{\jmp}[3]{{\it  J.\ Math.\ Phys.\ }{{ \bf #1} {(#2)} {#3}}}
\nc{\jpa}[3]{{\it  J.\ Phys.\ A\ }{{\bf #1} {(#2)} {#3}}}
\nc{\jpc}[3]{{\it  J.\ Phys.\ C\ }{{\bf #1} {(#2)} {#3}}}
\nc{\jap}[3]{{\it J.\ Appl.\ Phys.\ }{{\bf #1} {(#2)} {#3}}}
\nc{\jpsj}[3]{{\it J.\ Phys.\ Soc.\ Japan\ }{{\bf #1} {(#2)} {#3}}}
\nc{\lmp}[3]{{\it Lett.\ Math.\ Phys.\ }{{\bf #1} {(#2)} {#3}}}
\nc{\mpl}[3]{{\it  Mod.\ Phys.\ Lett.\ }{{\bf #1} {(#2)} {#3}}}
\nc{\ncim}[3]{{\it  Nuov.\ Cim.\ }{{\bf #1} {(#2)} {#3}}}
\nc{\np}[3]{{\it  Nucl.\ Phys.\ }{{\bf #1} {(#2)} {#3}}}
\nc{\pr}[3]{{\it Phys.\ Rev.\ }{{\bf #1} {(#2)} {#3}}}
\nc{\pra}[3]{{\it  Phys.\ Rev.\ A\ }{{\bf #1} {(#2)} {#3}}}
\nc{\prb}[3]{{\it  Phys.\ Rev.\ B\ }{{{\bf #1} {(#2)} {#3}}}}
\nc{\prc}[3]{{\it  Phys.\ Rev.\ C\ }{{\bf #1} {(#2)} {#3}}}
\nc{\prd}[3]{{\it  Phys.\ Rev.\ D\ }{{\bf #1} {(#2)} {#3}}}
\nc{\prl}[3]{{\it Phys.\ Rev.\ Lett.\ }{{\bf #1} {(#2)} {#3}}}
\nc{\pl}[3]{{\it  Phys.\ Lett.\ }{{\bf #1} {(#2)} {#3}}}
\nc{\prep}[3]{{\it Phys\. Rep.\ }{{\bf #1} {(#2)} {#3}}}
\nc{\prsl}[3]{{\it Proc.\ R.\ Soc.\ London\ }{{\bf #1} {(#2)} {#3}}}
\nc{\ptp}[3]{{\it  Prog.\ Theor.\ Phys.\ }{{\bf #1} {(#2)} {#3}}}
\nc{\ptps}[3]{{\it  Prog\ Theor.\ Phys.\ suppl.\ }{{\bf #1} {(#2)} {#3}}}
\nc{\physa}[3]{{\it  Physica\ A\ }{{\bf #1} {(#2)} {#3}}}
\nc{\physb}[3]{{\it  Physica\ B\ }{{\bf #1} {(#2)} {#3}}}
\nc{\phys}[3]{{\it Physica\ }{{\bf #1} {(#2)} {#3}}}
\nc{\rmp}[3]{{\it  Rev.\ Mod.\ Phys.\ }{{\bf #1} {(#2)} {#3}}}
\nc{\rpp}[3]{{\it Rep.\ Prog.\ Phys.\ }{{\bf #1} {(#2)} {#3}}}
\nc{\sjnp}[3]{{\it Sov.\ J.\ Nucl.\ Phys.\ }{{\bf #1} {(#2)} {#3}}}
\nc{\spjetp}[3]{{\it Sov.\ Phys.\ JETP\ }{{\bf #1} {(#2)} {#3}}}
\nc{\yf}[3]{{\it Yad.\ Fiz.\ }{{\bf #1} {(#2)} {#3}}}
\nc{\zetp}[3]{{\it Zh.\ Eksp.\ Teor.\ Fiz.\  }{{\bf #1}  {(#2)} {#3}}}
\nc{\zp}[3]{{\it Z.\ Phys.\ }{{\bf #1} {(#2)} {#3}}}
\nc{\ibid}[3]{{\sl ibid.\ }{{\bf #1} {#2} {#3}}}
\nc{\rf}[1]{(\ref{#1})}
\nc{\nn}{\nonumber \\*}
\nc{\bfB}{\bf{B}}
\nc{\bfv}{\bf{v}}
\nc{\bfx}{\bf{x}}
\nc{\bfy}{\bf{y}}
\nc{\vx}{\vec{x}}
\nc{\vy}{\vec{y}}
\nc{\oB}{\overline{B}}
\nc{\oI}{\overline{I}}
\nc{\oR}{\overline{R}}
\nc{\rar}{\rightarrow}
\nc{\ti}{\times}
\nc{\slsh}{\hskip-5pt/}
\nc{\sm}{Standard~Model~}
\nc{\MP}{M_{\rm Pl}}
\nc{\tp}{t_{\rm Pl}}
\nc{\ave}{\bar{E}}

\nc{\eff}{{\rm eff}}
\nc{\kk}{\vek{k}}
\nc{\pp}{{\rm p}}
\nc{\ga}{g_{a\gamma}}
\nc{\vv}{\\}
\nc{\eee}{{\bf E}}
\nc{\bbb}{{\bf B}}
\nc{\qcd}{T_{\rm QCD}}
\nc{\G}{\rm G}
\def\vec#1{{\bf #1}}
\begin{document}
{\title{\vskip-2truecm{\hfill {{\small \\
        }}\vskip 1truecm}
{\bf Primordial Magnetic Fields}}
{\author{
{\sc  Kari Enqvist$^{1}$}\\
{\sl\small Department of Physics and Helsinki Institute of Physics}\\
{\sl\small P.O. Box 9, FIN-00014 University of Helsinki,
Finland}
}
\maketitle
\vspace{1cm}
\begin{abstract}
\noindent
The explanation of the observed galactic magnetic fields may require
the existence of a primordial magnetic field. Such a field may arise
during the early cosmological phase transitions, or because of
 other particle physics
related phenomena in the very early universe reviewed here.
The turbulent 
evolution of the initial, randomly fluctuating
microscopic field to a large-scale macroscopic  field can be described
in terms of a shell model, which provides an approximation
to the complete magnetohydrodynamics. The results indicate that there is an
inverse cascade of magnetic energy whereby the coherence of the magnetic
field is increased by many orders of magnitude. Cosmological seed fields 
roughly of the order of $10^{-20}$ G at the scale of protogalaxy, 
as required by the dynamo explanation
of galactic magnetic fields, thus seem plausible.
\end{abstract}
\vfil
\footnoterule
{\small $^1$enqvist@pcu.helsinki.fi}

\thispagestyle{empty}
\newpage
\setcounter{page}{1}


\section{Introduction}
Apart from the baryon number and the spectrum of energy density fluctuations,
the physical processes that took place in the very early universe do not 
have many consequences that could  still be directly detectable today.
Most observables have been washed away by the thermal bath of the
pre-recombination era. One possibility, which has recently received
increased attention, is offered by the large-scale magnetic fields 
observed in a number of galaxies, in galactic halos, and in clusters of 
galaxies  \cite{observe,becketal}. 
The astrophysical mechanism
responsible for the origin of the galactic magnetic fields is not 
understood. Usually one postulates a small seed field, which can then be either
enhanced by the compression of the protogalaxy, 
and/or exponentially amplified by the
turbulent fluid motion as in the dynamo theory  \cite{dynamo}. The
exciting possibility is that the seed field could be truly 
primordial \cite{kulsrud},
in which case cosmic magnetic fields could provide direct information
about the very early universe. 

Early magnetic fields could then play an important role in
particle cosmology by modifying the dispersion or clustering properties
of various particles.  One particular example is the fate of the neutrino: 
because of their magnetic moments, Dirac
neutrinos propagating in the background of a magnetic field would be
subject to a  spin flip \cite{eers},
so that a left-handed neutrino can be turned into a right-handed neutrino,
giving rise to an extra effective neutrino degree of freedom and thereby
affecting primordial nucleosynthesis. Dark matter particles 
could also be sensitive to the presence of a magnetic field. For instance,
axions couple to magnetic fields, but perhaps surprisingly, it can 
be shown that despite the coupling, cold
axion oscillations are not much affected by the presence of a primordial
magnetic field \cite{jarkkoax}.

The issue at hand is then: is it possible that primordial magnetic
fields of significant strength exist? To answer this,  first one has to
find a mechanism in the early  universe which is able to produce a
 large enough magnetic field.  There are various proposals, a number 
of which are
based on the early cosmological phase transitions, which are discussed
in Sect 3. The second problem is to explain how the initial field, 
which is expected to be random as it is created  by microphysics and
having correlation lengths typical to microphysics, can grow up to be
coherent enough at large length scales. This is a problem in
magnetohydrodynamics which is discussed in Sect 4.
%
\section{Observation and the origin of galactic fields}
\subsection{Observing cosmic magnetic fields}
Cosmic magnetic fields can be observed
indirectly both at optical and radio wavelengths (for recent reviews,
see
\cite{observe,becketal}). 
Electrons moving in a magnetic field emit synchrotron radiation,
and both its intensity and polarization can be used to extract
information about the magnetic field; the radio emission is roughly
proportional to $n_eB^2$, where $n_e$ is the electron density.
 However, one first needs to fix
 the relative magnitudes of the electron and
magnetic field densities. Usually equipartition of 
magnetic and plasma energies is assumed, but in some cases free 
electron densities can be determined independently. 

Information about distant magnetic fields, e.g. in clusters of
galaxies, has been obtained by studying the Faraday rotation of polarized
light.
The method is based on the fact that
the plane of polarization of linearly polarized electromagnetic wave
rotates as it passes through plasma supporting a magnetic field. The
rotation angle $\Delta \chi$ depends on the strength and extension of the 
magnetic field,
the density of plasma, and on the wavelength $\lambda$ of radiation.
The Faraday rotation measure (RM) out to a maximum redshift $z_m$
is defined as
\be{rotm}
RM(z_m)\equiv
{\Delta \chi \over \Delta \lambda^2} = 8.1\times 10^5\int n_eB_{l}(z)
(1+z)^{-2}dl(z)
~{\rm rad\;m^{-2}}~,
\ee
where $B_{l}(z)$ is the strength of the magnetic field along the line of sight
in Gauss,
$n_e(z)$ is in cm$^{-3}$,
and
\be{dl}
dl(z)=10^{-6}H_0^{-1}(1+z)(1+\Omega z)^{-1/2}dz\;{\rm Mpc}
\ee
for a $\Lambda=0$ Friedmann universe, and $H_0$ is in kms$^{-1}$Mpc$^{-1}$
This method requires some independent information about the electron
density along the line of sight and the field reversal scale. In our galaxy
this may be deduced from the relative pulse delay and frequency of
pulsars, but pulsars are too faint to be observed in external galaxies.
Sometimes $n_e$ can be measured independently through X-ray 
bremsstrahlung, or by absorption line transitions in the object to be
measured.

The observed RM is sensitive mainly to
a regular magnetic field as small scale fluctuations tend to cancel out
in the integral \eq{rotm}. The more tangled the field, the smaller RM
is. To determine RM accurately one needs to measure at least three 
different wavelengths because the orientation of the polarization plane is
ambiguous by multiples of $\pi$. The sign of RM allows to distinguish between 
the two opposite directions of the regular field. 
Magnetic field irregularities
may be detected through depolarization, which tends to wipe out 
Faraday rotation at small wavelengths.

The 
Zeeman splitting of the spectral lines would provide a direct
measure of the strength of the magnetic field, but the shifts
are very small and this method is 
applicable mainly to our own galaxy. It has been used at radio frequencies
to study various galactic molecular clouds.

Magnetic fields at the level of few $\mu{\rm G}$
have been detected in galaxies, in galactic halos, and
in clusters of galaxies. Magnetic field appears to be a normal feature
of spiral galaxies, where their mean field strengths range from 
$4\mu{\rm G}$ in M33 to about $12\mu{\rm G}$ \cite{becketal}. 
Somewhat larger values can
be found in the spiral arms. The field in spiral galaxies
shows often either an axisymmetric or bisymmetric form following the spiral
structure. It also seems that the magnetic
field is not a relative newcomer but existed already some $6\times 10^9$
years ago, as witnessed by the Faraday rotation observed in a high
(z=0.395) redshift galaxy \cite{kpz}. 

Edge-on galaxies  often show vertical
dust lines, indicating the existence of vertical magnetic field lines
escaping from the disc. Synchrotron radiation emitting halos confirm
this picture. The halo magnetic fields are associated with the outward flow
of gas and dust (``galactic wind'') generated by 
stellar activity and supernovae in the disc \cite{observe}.

Faraday rotation measurements of
radio sources inside and behind galactic clusters, combined with electron 
density estimates obtained from the X-ray spectrum, indicate that  also many
of them have sizeable magnetic fields. An analysis of 
the RM distribution of radio sources seen through or near 
50 Abell clusters
\cite{ktk} implies a field strength of about 
$2\mu\G\; (L_{frs}/10\ {\rm kpc})^{-1/2}h_{50}^{-1}$, 
where $L_{\rm frs}$ is the
typical field reversal scale and $h_{50}$ is the Hubble parameter
in units of 50 kms$^{-1}$Mpc$^{-1}$; there are indications that 
$L_{\rm frs}$ in 
clusters is
actually rather small, of the order of kpc. 
For example, a recent
polarization data of a radio galaxy belonging to the
Coma cluster implies \cite{feretti} the existence of a core 
magnetic field with 
$B\gsim 8.3h^{1/2}_{100}\mu\G$ tangled at scales less than 1 kpc.
The fact that clusters show
a magnetic field which has a strength comparable with the 
much denser interstellar medium of galaxies is somewhat surprising,
but would of course be natural if the field had a truly primordial 
origin. Perhaps one piece of evidence pointing towards the 
existence of such a cosmological field is provided by the Hydra A
cluster, for which the RMs have been determined continuously from within
few arcseconds of the radio core out to a distance of 40'', corresponding
to 55 kpc.
The high Faraday rotations observed \cite{hydra}, which show
strong asymmetry between the north and south lobes of the cluster, can be
explained in terms of a $6\ \mu{\rm G}$ field coherent over the scale of
100 kpc, together with a tangled component of $\sim 30\ \mu{\rm G}$.

The currently favoured explanation for the origin of the large scale 
galactic magnetic fields is the $\alpha-\omega$ dynamo \cite{dynamo},
which through turbulence and differential rotation amplifies a small
frozen-in
seed field ${\bf B}_0$ to the observed $\mu{\rm G}$ field. 
An initially toroidal 
seed field, which is carried along on the disc of a rotating (spiral) galaxy,
is locally distorted into a loop by the up- or downward stochastic drift of the
gas. As the gas moves away from the plane of the disc, the pressure
decreases and the gas expands; at the same time it is subject to a Coriolis 
force which will rotate it. The magnetic field lines, glued to the gas,
will follow and thus a poloidal component perpendicular to ${\bf B}_0$
is generated. The small poloidal loops will reconnect and coalesce
to produce a large scale field. This is
the so-called $\alpha$-effect. Because the disc does not rotate like a 
rigid body, the field lines will be wrapped and the poloidal field
will induce a toroidal field; this is the $\omega$-effect. 
Thus the dynamo mechanism does not only produce a large scale
field but can also, to some extent, predict the shape of the field.

The dynamo saturates when the growth enters the non-linear regime.
A typical growth time is of the order of $10^9$ years, with the
rotation period of the galaxy setting a lower limit on the growth
time. It should however be pointed out that there are some indications that
the saturation might actually 
be too fast for a large-scale field to form  \cite{dproblems}.
In any cae, the strength of the required initial seed field is rather
uncertain, but as a rule of thumb one could use a value like
$10^{-20}$ G on a comoving scale of the protogalaxy (100 kpc).

Another possibility is that the galactic field results
directly from a primordial field, which gets compressed when
the protogalactic cloud collapses. The primordial field strengths
needed are however quite large, of the order of $10^{-9}-10^{-10}$ G.
In any case, it appears as if a primordial field is required to
explain the observed galactic magnetic fields,
although it is conceivable that a purely astrophysical solution could 
exist as well. For instance,  within the nucleus of  
the ``starburst galaxy'' M82 there is a
small but extremely active zone of star formation, which gives rise
to considerable outflow of material. Associated with it is an outward
trasportation of magnetic field of about 50 $\mu{\rm G}$, 
which manifests itself through field
lines connected with the streaming motions of the synchrotron emitting
relativistic plasma
\cite{becketal, m82}. The field structure cannot be 
attributed to the action of an $\alpha-\omega$ dynamo.
This demonstrates that other mechanisms exist which
are able to amplify and create magnetic fields at least in the halos
of galaxies. Whether purely astrophysical processes could explain
galactic magnetic fields remains however to be seen, although
some suggestions have been put forward. 
For instance, density and temperature fluctuations could create weak
magnetic fields via the so-called Bierman battery mechanism which
subsequently could be enhanced by turbulence \cite{battery}. 
A more exotic suggestion is that 
rotating black holes could be responsible for the seed
field
\cite{vilenkin}.  The idea requires asymmetric particle emission from
the black hole to create a current and hence a magnetic field.

\subsection{Magnetohydrodynamics in curved space}
Before discussing limits on primordial magnetic fields, it is useful
to have a brief theoretical digression.
The behaviour of the early magnetic field is described by the
magnetohydrodynamic (MHD) equations \cite{relMHD}. To derive them, one assumes
that the  energy-momentum tensor is the sum of an ideal fluid and 
Maxwell parts which, 
in the absence of viscosity and magnetic diffusivity, reads
\be{teemunu}
T^{\mu\nu}=(p+\rho) U^\mu U^\nu + p g^{\mu\nu} 
+{1\over4\pi}\left(F^{\mu\sigma} {F^\nu}_\sigma
-{1\over4}g^{\mu\nu}F_{\lambda\sigma}F^{\lambda\sigma}\right)~.
\ee
Here $U^\mu$ is the four-velocity of the plasma,
normalized as $U^\mu U_\mu=-1$,  and $F_{\mu\nu}=
\partial_\mu A_\nu-\partial_\nu A_\mu$ is the electromagnetic field tensor.
Note that, as long as diffusion can be neglected, the presence of the magnetic
field does not change the equation of state of an ideal fluid. 
The equations of motion for the fluid arise from energy-momentum conservation 
\begin{equation}
{T^{\mu\nu}}_{;\nu}\equiv{1\over\sqrt{-g}}{\partial\over\partial x^\nu}
\sqrt{-g}{T^{\mu\nu}}+\Gamma^\mu_{\nu\lambda}{T^{\nu\lambda}}=0~.
\label{tmunu}
\end{equation}
The Maxwell equations read
\begin{equation}
{F^{\mu\nu}}_{;\nu}= J^\mu,\quad
F_{[\mu\nu,\lambda]}=0.
\label{maxwell}
\end{equation}
In a flat, isotropic and homogeneous 
universe with a Robertson-Walker metric
$ds^2=-dt^2+R^2(t) d{\bf x}^2$,
we may define $F_{\mu\nu}$ in terms of the electric and magnetic fields as
\begin{equation}
F_{i0}=RE^i,\quad F_{ij}=\epsilon_{ijk}R^2B^k,
\label{ejab}
\end{equation}
where latin letters go from 1 to 3. With this definition the expression for
the total energy has no $R$-factors and takes therefore the familiar form
\begin{equation}
T^{00}=(p+\rho)\gamma^2 - p +{1\over 2}({\bf B}^2+{\bf E}^2),
\label{T00}
\end{equation}
where $\gamma=U^0$. 

The 
MHD equations take a particularly simple form when using the 
conformal metric $ds^2=a^2(\tau)(d\tau^2-d\vec x^2)$.
The Maxwell equations can then be written  as
\be{max1}
{\partial \tilde{{\bf B}}\over\partial\tilde{t}}=
-\nabla\times\tilde{\bf E},\quad\mbox{\boldmath $\nabla$}
\cdot\tilde{{\bf B}}=0,
\ee
and
\be{max2}
\tilde{\bf J}=\mbox{\boldmath $\nabla$}\times\tilde{{\bf B}}
-{\partial \tilde{\bf E}\over\partial \tilde{t}},
\quad\mbox{\boldmath $\nabla$}\cdot\tilde{\bf E}=\tilde\rho_e
\ee
where the scaled variables are defined as 
\be{tildet}
\tilde{\rho}_e=a^4\rho_e~,
\quad\tilde{{\bf B}}=a^2{\bf B}~,\quad
\tilde{\bf J}=a^3{\bf J}~,\quad{\rm and}\quad\tilde{\bf E}=a^2{\bf E}.
\ee
Here $\rho_e$ is the charge density. 

The MHD equations in curved space, using conformal metric, 
thus have a similar form to those in flat
space. Let us therefore consider the flat space for simplicity.
The evolution of the magnetic field is described by
\be{magf}
{\partial\vec B\over \partial t}=\nabla\times(\vec v \times \vec B)
-\nabla\times\frac 1\sigma(\nabla\times\vec B)~,
\ee
which follows from the Maxwell equations \eqs{max1}{max2}
and Ohm's law.
In a perfect conductor  
$\sigma=\infty$ and field lines are frozen in to the fluid. If  $\sigma$
is finite, the fluid can slip through the field lines. 
The quantitative measure of the freezing is given by the magnetic Reynolds
number $R_M$. It can be defined through 
the ratio of the two terms in \eq{magf} as
\be{reynold}
R_M\simeq {\sigma \vert\nabla\times(\vec v \times \vec B)\vert \over
\vert\nabla^2\vec B\vert}\sim \sigma v L~,
\ee
where $L$ is the length scale and $v$ is the bulk or drift velocity of 
the fluid. If $R_M\gg 1$, the magnetic flux
lines will be frozen into the plasma.
In the
astrophysical context the fluid can often be considered as
an extremely good conductor with $R_M\gg 1$. The early universe is also
a good conductor \cite{turnerwidrow,cond} with $\sigma \sim T$ and 
$L\sim M_{Pl}/T^2$ so that easily $R_M\gg 10^{10}$. Large $R_M$
signals the onset of turbulence, which thus could be expected
to play a role in the early universe magnetohydrodynamics, as will
be discussed in Sect. 4.

\subsection{Cosmologial fields: limits and observations}
Observing cosmological, intergalactic magnetic fields would naturally 
be of great
importance as it would strengthen the case for their truly primordial
origin. While the fields in the neighbourhood of galaxies or clusters
of galaxies could conceivably originate in the ejected magnetoplasma,
magnetic fields in the voids, which are devoid of baryonic matter,
would be much harder to explain in terms of conventional astrophysics.

Energetic $\gamma$-rays can traverse the intergalactic voids, and
it has been claimed by Plaga \cite{plaga}
that the arrival times of extragalactic
$\gamma$-rays could be used to detect fields as weak as $10^{-24}$ G.
The idea is that 
cosmic rays, originating from far-away objects such as
QSOs or gamma-ray bursters, 
scatter  off the background cosmic magnetic field. This gives rise to pair
production and a delayed $\gamma$-ray which could then be observed, and
the ratio of prompt to delayed $\gamma$'s provides a measure of the
strength of the intergalactic magnetic field. The energy 
spectrum of $\gamma$-rays
is likewise affected by the intergalactic magnetic field \cite{los}.
Ultra-high cosmic ray protons 
would also be deflected by the intergalactic magnetic
fields so that their arrival times could be used \cite{cray} 
to set bounds as low
as $10^{-12}\G$.

A constraint on the strength of a cosmological magnetic field
has been obtained by considering the rotation measure of a sample of 
309 
galaxies and quasars, which yields the limit \cite{vallee} $B \le 10^{-9} \G
\; (\Omega_{IG}h_{100}/0.01)^{-1}$, where $\Omega_{IG}$ is the fraction of
the ionized gas density of the critical density in the intergalactic medium.
 The assumption here is that the field is
coherent over the horizon scale, and the limit is weakened if the field
is tangled at smaller scales, as in fact would seem likely.
It has also been suggested that 
the power spectrum of the cosmological
field could be determined by studying the correlations in the RM
for extragalactic sources  \cite{kolatt}.

If a primordial magnetic field is present at the time of recombination,
it will give rise to anisotropic pressures. These would distort the
microwave background. Barrow, Ferreira and Silk \cite{cmblimit}
have recently considered the effect of a magnetic field on the evolution
of shear anisotropy $\sigma_A$ in a general anistropic flat
universe. For a fluid with an equation of state given by $p=(\gamma-1)\rho$
(with $0<\gamma\le 4/3$) it was found that
\be{bfs3}
{\sigma_A\over H}={4\over 2-\gamma}\left({\rho_B\over\rho}+{\rho_{ga}\over\rho}
\right)+\delta t^{(\gamma-2)/\gamma}~,
\ee
where $\rho_{ga}$ is the gas anistropy, $\rho_B=B^2/8\pi$ is the magnetic 
energy density, and $\delta$ is a constant. The angular anisotropy 
$\Delta T$ is then
directly proportional to $\sigma_A/H$.
Barrow, Ferreira and Silk then compared the result with the 4-year COBE data
set, and assuming that the whole observed anisotropy is due to
magnetic stresses only, concluded that, 
if the field can be taken homogeneous, 
$B \le 3.4\times 10^{-9} \G
\; (\Omega_{0}h_{50}^2)^{1/2}$. 
This is a more stringent limit
than what can be obtained  \cite{cost,ksv,dariohector} 
from primordial nucleosynthesis
considerations. 

The anisotropy limit also indicates
that the  distortions of the Doppler peaks of the microwave 
background \cite{adamsetal} due to a homogeneous magnetic field 
are unobservable, although distortions due to an inhomogeneous 
magnetic field may still be observable. The distortions arise because
in the presence of a magnetic field the 
wave patterns generating fluctuations in the energy density change. There
are magnetosonic waves, which in the absence of a magnetic field would
correspond to sound waves, together with Alfv\'en waves
propagating with velocity $v_A$, which however 
do not induce energy density fluctuations but affect the velocities
of baryons and photons \cite{adamsetal} . 
The fast magnetosonic waves  effectively
change the sound velocity $v_S$ in the baryon fluid  by $v_S^2\to
v_S^2+v_A^2\sin^2\theta$, where $\theta$ is the angle between 
$\vec B$ and the wave vector $\vec k$, thereby changing the Doppler
peak structure.

The Faraday rotation in the  polarization
of the microwave background could still be observable, as has been
argued by Kosowsky and Loeb 
\cite{kosowsky}. They estimate that for a primordial field corresponding
to the present value of $10^{-9} \G$, coherent on the scale of the last
scattering surface, there should be a rotation angle
of $1^{o}$ at 30 GHz. Such a signal requires a beam size of order 
$10'$ and could conceivably be detected in the future
microwave background maps.

\subsection{Limits from primordial nucleosynthesis}
Primordial nucleosynthesis offers a laboratory for many particle
physics ideas. It also provides a testing gorund for  
 primordial  magnetic fields, which
would modify the calculated
element abundances through two effects \cite{cost, dariohector}. 
The extra magnetic field energy density
would accelerate the expansion rate of the universe, and hence increase the
predicted abundance of $^4{\rm He}$. If one defines the magnetic
field at the end of nucleosynthesis ($T=10^9~ K$) in terms of the critical
field $B_c\equiv eB/m_e^2= 4.4\times 10^{13}\G$ with $B=\gamma B_c$,
one finds numerically \cite{dariohector}
that $\gamma=3\times 10^{-3}$ is sufficient to push 
the relative $^4{\rm He}$-abundance
above 0.24. There is also a small effect on
the $^7{\rm Li}$ and $({\rm D+^3He})$ abundances, but these are negligible.

However, magnetic fields affect the Hubble rate yet in another, more 
subtle way \cite{cost,dariohector}. 
The statistical distributions of relativistic electrons and
positrons change because of the background magnetic field. The particles
are no longer plane waves but occupy Landau levels, with the dispersion
relation given by (for $B$ along z-axis)
\be{edispersion}
E=\left(p_z^2+m_e^2+2eBn_s\right)^{1/2}+m_e\kappa~,
\ee
where $n_s=n+1/2-s_z$ with $s_z=\pm 1/2$ and $n$ the Landau level,
and $\kappa$ is the anomalous magnetic moment term, which for
fields less than about $10^{16}\G$ reads $\kappa=e\alpha B/(4\pi m_e^2)$;
for subcritical fields, its effect is small. The number density of states
in the interval $dp_z$ is then
\be{nmdensi}
(2-\delta_{n_s,0}){eB\over (2\pi)^2}dp_z~.
\ee
 If $eB\ll T^2$, as
is natural, a fraction of the electron-positron pairs condense in the
lowest Landau level. As a consequence, the number and energy densities
of electrons and positrons increase with respect to the zero field
case.

A primodial magnetic field affects primordial nucleosynthesis
also 
by increasing $n\leftrightarrow p$ reaction rates \cite{cost,dariohector}. 
Hence 
$^4{\rm He}$ abundance tends to decrease, partly counteracting the effect
due to change in 
the Hubble expansion rate. It turns out, however, that this effect
is also small. More important is the modification of the phase space of the 
electrons: increasing number and energy densities causes a decrease
in all of the weak interaction rates. 
Taking all these effects into account and requiring that
the $^4{\rm He}$ abundace should not exceed 0.245, Grasso and Rubinstein 
obtain \cite{dariohector}  the upper limit 
\be{nsgr}
B\le 1\times 10^{11}\G~
\ee
at $T=10^9~ K$. A similar conclusion has also been obtained in \cite{cost}.
Here the magnetic field was assumed to be uniform; if inhomogeneities 
on the scale much less than the horizon are
present, as seems likely, the effect of the magnetic field on weak
rates could be averaged out. In this case Grasso and Rubinstein obtain
the limit
\be{nsgr2}
B\le 1\times 10^{12}\G~.
\ee
This assumes a magnetic field coherence length $L_0$ in the range
$10\ll L_0\ll 10^{11}$ cm.
\section{Generating primordial magnetic fields}
\subsection{Early phase transitions}
There are a number of proposals for the origin of  magnetic fields
in the early universe.
Fluctuations in the electromagnetic field in a relativistic plasma
are by themselves sufficient for generating a small scale random magnetic
field \cite{lemoine}. To obtain a seed field for galactic magnetic fields,
one however needs fields with much larger coherence lengths.
A natural place to look for large scale magnetic fields would be
the early cosmological phase transitions, such as the electroweak or
the QCD phase transition.

Magnetic fields may arise in a electrically neutral plasma if
local charge separation happens to take place, thus creating a local 
current. It has been proposed that this could occur
during a first order QCD \cite{qcdsep} or EW \cite{ewsep,siglolinto}
phase transition, which proceed by nucleating bubbles of
the new phase in the background of the old phase. 
There one finds net baryon number gradients 
at the phase boundaries, providing the basis for charge separation, 
and the seed fields arise through 
instabilities in the fluid flow. For that one has to require
that the growing modes are not damped.
Turbulent flow near the walls of the bubbles is then expected to amplify
and freeze the transient seed field. 
The various hydrodynamical
features have been studied 
in linear perturbation theory by Sigl, Olinto and Jedamzik \cite{siglolinto}, 
who argued that 
on a 10 Mpc comoving scale, field strengths of the order of
$10^{-29}$ G for EW and $10^{-20}$ G for QCD could be obtained.
These might further be enhanced by several orders of magnitude 
by hydromagnetic turbulence \cite{beo1,beo2},
which will be discussed in Section 4.3. 
\subsection{Bubble collisions}
In a first order phase transition the phases of the complex order parameter
$\Phi = \rho e^{i\Theta}/\sqrt{2}$ of the nucleated bubbles are 
uncorrelated. When
the bubbles collide, there arises a phase gradient which acts as
a source for gauge fields. The phase itself is not a gauge invariant
concept, but it has been pointed out by Kibble and Vilenkin \cite{kv}
that a gauge invariant phase difference 
can be defined in terms of an
integral over the gradient $D_\mu\Theta$.

Magnetic field generation in the collision of phase transition bubbles
has been considered in the abelian Higgs model \cite{kv,ae3}. One  assumes that 
inside the bubble the radial mode
$\rho$ settles rapidly to its equilibrium value $\eta$ and can thus be treated
as a constant. The dynamical variables are thus $\Theta$ and the gauge
field $A^\mu$.
The starting point is the U(1)-symmetric lagrangian
\be{wuwwvvxx}
\cL=-\frac{1}{4}F_{\mu\nu}F^{\mu\nu}+D_{\mu}\Phi (D^{\mu}\Phi )^{\dagger}+
V(\vert\Phi\vert ),
\ee
where the potential
$V$ is assumed to have minima at $\rho=0$ and $\rho=\eta$.
The simplest case is that two spherical bubbles nucleate, 
one at $(x,y,z,t)=(0,0,z_0,0)$ and the other at $(x,y,z,t)=(0,0,-z_0,t_{0})$,
and keep expanding with the velocity $v$ even after colliding. Because of the 
symmetry of the problem, the solutions to the equations of motion are
functions of $z$ and $\tau=\sqrt{t^2-x^2-y^2}$ only, and the U(1) gauge field
$A^{\mu}=x^{\mu}f(\tau ,z)$. One finds \cite{kv} that
the generated magnetic field is rapidly oscillating and orthogonal to
the z-axis and in this case confined inside the intersection region.
It  has a ring-like shape in the
(x,y)-plane.
It can also be shown that subsequent collisions of the bubbles, which now may 
have a magnetic field inside the bubbles,  nevertheless lead to
a qualitatively similar outcome  \cite{ae3}. 

There are however two additional important ingredients which need to be
taken into account: the high but finite conductivity \cite{turnerwidrow,cond}
of the primordial plasma
  and the fact that in the electroweak
phase transition the bubbles will in fact intersect with non-relativistic
velocities  \cite{velocity}. Finite conductivity gives rise to diffusion,
the consequence of which
is to smooth out the rapid oscillations
of $\vec B$, whereas low $v$ will permit the magnetic flux to escape
the intersection region and penetrate the colliding bubbles, where its
evolution will
be governed by usual magnetohydrodynamics. 

The strength of the generated magnetic field depends on the bubble
wall velocity in an essential way  \cite{ae3}.
In the electroweak case the initial growth of the
bubble wall is by subsonic deflagration, with velocities of the order 
of 0.05$c$,
depending on the assumed friction strength
  \cite{velocity}. The wall is preceded by a shock
front, which may collide with the other bubbles. This results in reheating, and
oscillations of the bubble radii, but
eventually a phase equilibrium is attained. The ensuing bubble growth is
very slow and takes place because of the expansion of the universe.
Because the universe has been reheated back to $T_c$,
no new bubbles are formed during the slow growth phase. 
Assuming that the abelian Higgs model results are applicable to the 
electroweak case, 
one may estimate that \cite{ae3}
\be{wuuutttt}
B \simeq 2.0\times 10^{20}
\sqrt{\gamma^2+2\gamma R}/R \, 
\G, 
\ee
where it has been assumed $\Theta_0=1$, $T_c=e\eta=100\GeV$;
the average distance between the nucleation 
centers 
$r_{ave}=9.5\times 10^{-8}t_H$ and the velocity $v=1.2\times 10^{-4}$
were taken as reference values.
Folding in the spectrum of separation of the adjacent shocked spherical 
bubbles  \cite{meyer}, averaging over all possible inclinations of
the ring-like magnetic field, and taking into account the
enhancement of magnetic energy due to an inverse cascade (see next Section),
one arrives at the estimate $B_{rms}\simeq 10^{-21} {\rm G}$ for 
the cosmological magnetic field at the scale of 10 Mpc today \cite{ae3}.

There are some indications that the behaviour of the abelian Higgs model
does not differ qualitatively from the full $SU(2)\times
U(1)$ case \cite{saffincope, dariotoni}. However, it has also been argued 
by T\"ornkvist \cite{ola} 
that in the Standard Model the non-abelian nature of the gauge fields
in fact forbids the formation of magnetic fields in two-bubble
collisions, which in effect can be gauge rotated away. 
The result is
valid for a specific intial bubble configuration and applies only for
the initial evolution. The question is thus still open, but it seems likely
that in bubble collisions magnetic fields are generated sooner or later.
For example, in a collision of three bubbles magnetic field can no
longer be rotated away \cite{ola}.
\subsection{Fluctuating Higgs gradients}
It has been pointed out by Vachaspati \cite{vach} that fluctuating Higgs field
gradients will induce a magnetic field. Such local fluctuations
are naturally present at the EW phase transition, no matter what
its
order, since the embedding of the electromagnetic field in  
$SU(2) \otimes U(1)_Y$ involves these gradients:
\be{xcxcqqq}
F_{ij}^{em}  =  - i(V^\dagger_i V_j - V_i V^\dagger_j)~,~~
V_i  =  \frac{2}{|\phi|} \sqrt{\frac{\sin \theta}{g}}~
\partial_i \phi ~~,                                           
\ee
where $\phi$ is the Higgs field. 
At the electroweak phase transition the field strength $F_{ij}^{em}$ 
can be expected to be constant and ${\cal O}(m_W)^2$ but
varies in a random way over larger distances.
The vector $V_i$ is also random, of course.
Its variation is due to the fact that the Higgs field $\phi$ makes a
random walk on the vacuum manifold of $\phi$.

To find out the root-mean-square field at large distance scales, one should 
make use of a proper statistical argument.
The question then is, what sort of an average should one use?
The line
average,
where the gradient vectors are taken to be the basic stochastic 
variables \cite{poul}, gives rise to a
scaling behavior $B\sim 1/\sqrt{L}$. Because measuring the strength
of $B$ by Faraday 
rotation of the emitted light involves the average along the
line of sight, this averaging method could indeed be appropriate in
these cases. It has however been argued that for the galactic dynamo,
only the volume average has dynamical significance \cite{hindmarsh}.
This has been shown
to imply that, in the absence of any turbulent enhancement, 
even under the most favourable circumstances the electroweak phase transition 
can manage only about $10^{-20}\G$ on the 1 Mpc scale \cite{hindmarsh}.

\subsection{Vacuum condensates}
Another, a more exotic possibility for generating primordial magnetic fields is
based on the observation that, due to quantum fluctuations,
the Yang--Mills vacuum is unstable in a large enough background magnetic
field \cite{savvidy} at zero temperature.

In the early universe the effective energy picks up thermal corrections
from the fermion, gauge boson, and Higgs boson loops. The detailed form of
the thermal correction depends on the actual model, but one may take the
cue from the SU(2) one--loop calculation, where they are
obtained by summing the Boltzmann factors $\exp(-\beta E_n)$ for the
oscillator modes
\be{wurrtmmm}
E_n^2=p^2+2gB(n+\frac 12)+2gBS_3+m^2(T), 
\ee
where $S_3=\pm 1/2~(\pm 1)$ for fermions (vectors bosons).
\eq{wurrtmmm} includes the thermally induced mass $m(T)\sim gT$,
 corresponding to
a ring summation of the relevant diagrams. Numerically, the effect of
the thermal mass turns out to be very important.
At high temperature, the leading behaviour is
given mainly by the bosonic contributions, and thus one arrives at 
the estimate \cite{ferro}
\be{zxccccvv}
\delta V_T^v=\frac {(gB)^2}{8\pi^2}\sum_{l=1}^{\infty}\int_0^{\infty}
\frac{dx}{x^3} e^{-K_l^b(x)}\left[\;
x{{\rm cosh}(2x)\over {\rm sinh}(x)}-1\right],  
\ee
where $K_l^a(x)={gBl^2/(4xT^2)}+{m_a^2x/(gB)}$ and numerically 
$\delta V_T^v\sim 0.02\;(gB)^2$ which serves only to shift the value of
$B$ at the vacuum slightly. Thus $B\ne 0$ is the state of the lowest energy
even at high temperature.

A local fluctuation will then trigger the creation of the
a new vacuum with non-zero non--abelian magnetic field 
inside a given particle horizon at scales $\mu \simeq T$.
The Maxwell magnetic field
is then just a projection of the non-abelian field.
The resulting magnetic field is exponentially suppressed  
and can safely be neglected in the case 
of the electroweak gauge group. 
 In the early
universe, however, where possibly a grand unified symmetry is
valid, the suppression may be less severe \cite{ferro}. It is also
attenuated by the running of the coupling constant.
One can then estimate that 
\be{zxcqqwwww}
B(T)=g_{\rm GUT}^{-1}\mu^2\exp \left(-{48\pi^2\over 11Ng^2}\right)
\left({T^2\over \mu^2}\right)\simeq
3\times 10^{42}\left({a(t_{\rm GUT})\over a(t)}\right)^2\;\G, 
\ee
where  the reference
number is for susy SU(5). The field is large enough to serve as the
seed field. 

Electroweak magnetic condensates have also  recently been considered by 
Cornwall \cite{Cornwall}, who has suggested that they would give rise
to magnetic fields with a net helicity
via the generation of electroweak Chern-Simons number. 
Joyce and Shaposnikov \cite{joyceshapo}
have argued that a
right-handed electron asymmetry, generated at the GUT scale,
could give rise to a hypercharge magnetic field $B_Y$ via
a Chern-Simons term. $B_Y$ is coupled to the baryonic current via
the anomaly, and would thus induce baryon number fluctations which
could survive until primordial nucleosynthesis \cite{shapogio}, making
it inhomogenous. These considerations deserve further study.

\subsection{Structure formation and and magnetic fields}
A natural idea is that primordial fields are induced by the mechanism
responsible for the structure formation, such as inflation. However,
the basic problem with inflation with regards to magnetic field generation 
is that the early universe is a good conductor so that, ignoring turbulence, 
the magnetic flux $\sim Ba^2$ 
tends to be conserved.
To avoid this, one needs to break the conformal
invariance somehow, as was first suggested by Turner and Widrow
\cite{turnerwidrow}, who considered couplings to the
curvature $R$ such as $RF^2$ and
$RA^2$, as well as photon-axion couplings. 
Dilaton coupling
of the form $e^\Phi F^2$ has also been considered \cite{ratra}, and interesting
field strengths can be obtained at the expense of tuning the dilaton 
coupling strength. If a phase transition takes place during the
inflationary period, a sufficiently large magnetic field could be created,
provided however that the phase transition takes place during the final 
5 e-foldings \cite{davisdimoinfl}.

Cosmic strings have also been suggested
to be responsible for magnetic fields \cite{strinwiggle}, 
the field being generated by 
the vortical motion inside the wakes of strings. There are however some 
open questions related to the stability of the strings, which might
be remedied if the strings are superconducting
\cite{dimostring}. There  also exist speculations \cite{sstrings} 
about superstringy origin
of the magnetic field.
\section{From microscopic to macroscopic fields}
\subsection{Magnetohydrodynamic turbulence}
Even assuming that a primordial magnetic field is created at some very
early epoch, a number of issues remain to be worked out before
one can say anything definite about the role primordial fields in the
formation of galactic magnetic fields. At the 
earliest times magnetic fields
are generated by particle physics processes with length scales typical to
particle physics. (It has been shown that such fields are stable against
thermal fluctuations \cite{martindavis}.)
The remaining question is whether it is at all possible
for the small scale fluctuations to grow to large scales, and what exactly is
the scaling behaviour of $B_{\rm rms}$ or the correlator
$\langle B(r+x) B(x)\rangle$.  To
study these problems one needs to consider the detailed evolution of the
magnetic field  to account for such issues as to what happens when uncorrelated
field regions come into contact with each other during the course of the
expansion of the universe. In general, turbulence is an essential
feature of such phenomena. These questions can only be answered by considering
MHD in an expanding universe  \cite{relMHD},
but a big problem is the very large Reynolds number of the early universe.
Realistic simulations are simply not possible. However, simulations 
can nevertheless be useful in giving an indication of the possible
trend. Starting with 
a random 2d magnetic field configuration in a radiation dominated 
FRW-universe, one finds that 
 the magnetic structures coalesce and there is a 
gradual emergence  of larger and larger magnetic scales  \cite{beo1}. 
Such a behaviour is encouraging,
indicating that the field is not really comovingly frozen.
On the other hand, because of the different topology, 
2d MHD is very different from the 3d MHD, and moreover,
the Reynolds number in the simulation (about 10) is still quite unrealistic.

Dimopoulos and Davis \cite{dimodavis} have also argued out that when two 
initially uncorrelated
domains come into contact, the field at the interface should untangle 
with the plasma bulk velocity $v$ to
avoid the creation of domain walls. They propose that the correlation
length  $\xi$ evolves according to
\be{zxcyyyy}
{d\xi\over dt}=H\xi + v~,
\ee
where $H$ is the Hubble parameter and 
the velocity $v$ depends dynamically on $B$ and should,
in principle, be determined from MHD. Qualitatively, this argument
also points towards a field not frozen in the plasma.
\subsection{Shell models}
Large $R_M$ raises the possibility of magnetohydrodynamic turbulence, or 
effects not visible in local perturbation theory.
In ordinary hydrodynamics many properties of
turbulence, in particular those related  to energy transfer and to the spectral
properties have been studied
successfully using a simple cascade model.
This is true not only qualitatively, but also quantitatively, which is
the reason why the cascade model is now much used in
nonlinear physics \cite{mogensbook}.

The basic idea  is that the interactions due to the nonlinear terms
in the MHD equations
are local in wavenumber space, and in $k$-space the quadratic nonlinear terms
become a convolution.
Interactions in $k$-space involving triangles with similar
side lengths have the largest contribution.
This has led to the shell model
which is formulated in
the space of the modulus of the wave numbers. This space is approximated by
N shells, where each shell consists of wave numbers with $2^n\leq k \leq
2^{n+1}$ (in the appropriate units). The Fourier transform of the velocity
over a length scale $k_n^{-1}$ ($k_n=2^n$) is given by the complex quantity
$v_n$, and $B_n$ denotes a similar quantity for the $B$-field.
Furthermore, the convolution is approximated by a sum over the nearest and
the next nearest neighbours,
\begin{equation}
N_n(v,B)=\sum_{i,j=-2}^2 C_{ij} v_{n+i} B_{n+j}.
\end{equation}
Here $v$ and $B$ have lost their vectorial character, which
reflects the fact that this model is not supposed to be an approximation
of the original equations, but should be considered as a toy model that
has similar {\it conservation} properties  as the original equations.

Velocity and magnetic fields are thus represented by scalars at the discrete
wave numbers $k_n=2^n$ ($n=1,...,N$), i.e. $k_n$ increases exponentially.
Therefore such a model can cover a large range of length scales
(typically up to ten orders of magnitude).
The important conserved quantity is $E_{\rm tot} a^4$, where
$E_{\rm tot}=\int T^{00}d^3x$ is the total energy. The 
resulting equations of motion
read, in conformal time $\tilde t$, \cite{beo1}
\begin{equation}
{\textstyle{4\over3}}\rho_0
{dv_n\over d\tilde{t}}=N_n(v,b),
\label{cascadeu}
\end{equation}
\begin{equation}
{db_n\over d\tilde{t}}=M_n(v,b),
\label{cascadeb}
\end{equation}
where $\rho_0$ is the background radiation density, 
and $\vec b =\vec Ba^2$.  The functions
$N_n$ and $M_n$ each have three terms (combinations of $v_j$ and $b_j$),
the form of which is determined by the MHD equations,
with three free parameters $A$, $B$, and $C$ \cite{beo1}. 
\subsection{Inverse cascade}
The numerical study of the cascade model requires of course that the
parameters $A,B,C$ are fixed so that the model has the same conservation
laws as the full-fledged MHD. One ignores the detailed evolution of
the density by setting $\rho \simeq \rho_0a^{-4}$ so that
\be{vali}
\int \left ( \frac43\rho_0\vec v^2+\frac12 \vec b^2\right)={\rm const.}
\ee
The model can then be solved 
numerically \cite{beo1}, and
the results
are shown in
Fig. 1, where the 
 transfer
of magnetic energy to larger and larger length scales is clearly seen. 
This process, the inverse cascade, is due to the
nonlinear terms giving rise to mode interactions. The initial magnetic
energy spectrum was chosen to be given by $E_M(t=0)\sim k$ with the 
total magnetic energy equal to $\rho_0$. The number of shells was $N=30$ 
so that length scales differing by ten orders of magnitude were covered.
\begin{figure}
\leavevmode
\centering
\vspace*{65mm}
\includegraphics{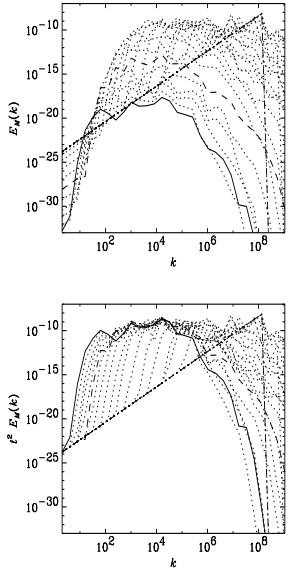}
\caption{
Spectra of the magnetic energy at different times.
The straight dotted-dashed line gives the initial condition ($t_0=1$),
the solid line gives the final time ($t=3\times10^4$), and the dotted
curves are for intermediate times (in uniform intervals of 
$\Delta\log(t-t_0)=0.6)$. 
}\label{pb_array}\end{figure}
It was
found that the integral scale, which measures where most magnetic energy
is concentrated and which is given by
\be{zxcjklklk}
l_0=\left.\int (2\pi/k)E_M(k)dk\right/\int E_M(k)dk,
\ee
where $E_M(k)$ is the magnetic energy spectrum,
increases with the Hubble time approximately like $t_H^{0.25}$.

However, around the time of
recombination the photon mean free path $\lambda_\gamma$ is very
large and photon diffusion becomes very efficient in smoothing out
virtually all inhomogeneities of the photon-baryon plasma \cite{silk}.
This process is often referred to as Silk damping, which corresponds to
a kinematic viscosity $\nu\simeq\lambda_\gamma$ (in natural units).
Silk damping may thus destroy the magnetic field,
as has been noted by Jedamzik, Katalinic and Olinto \cite{chicago}.
Therefore one has to follow numerically the evolution of the magnetic and kinetic
energy spectra in the presence of kinematic viscosity. 
The results \cite{beo2}  are presented in Fig. 2 and the
main point can be summarized as follows: in the cascade models
magnetic energy is transferred to large length scales even in the presence of 
large viscosity. Here the initial magnetic spectrum was chosen to be flat 
in accordance with the large time behaviour suggested by Fig. 1.
\begin{figure}
\leavevmode
\centering
\vspace*{90mm}
\includegraphics{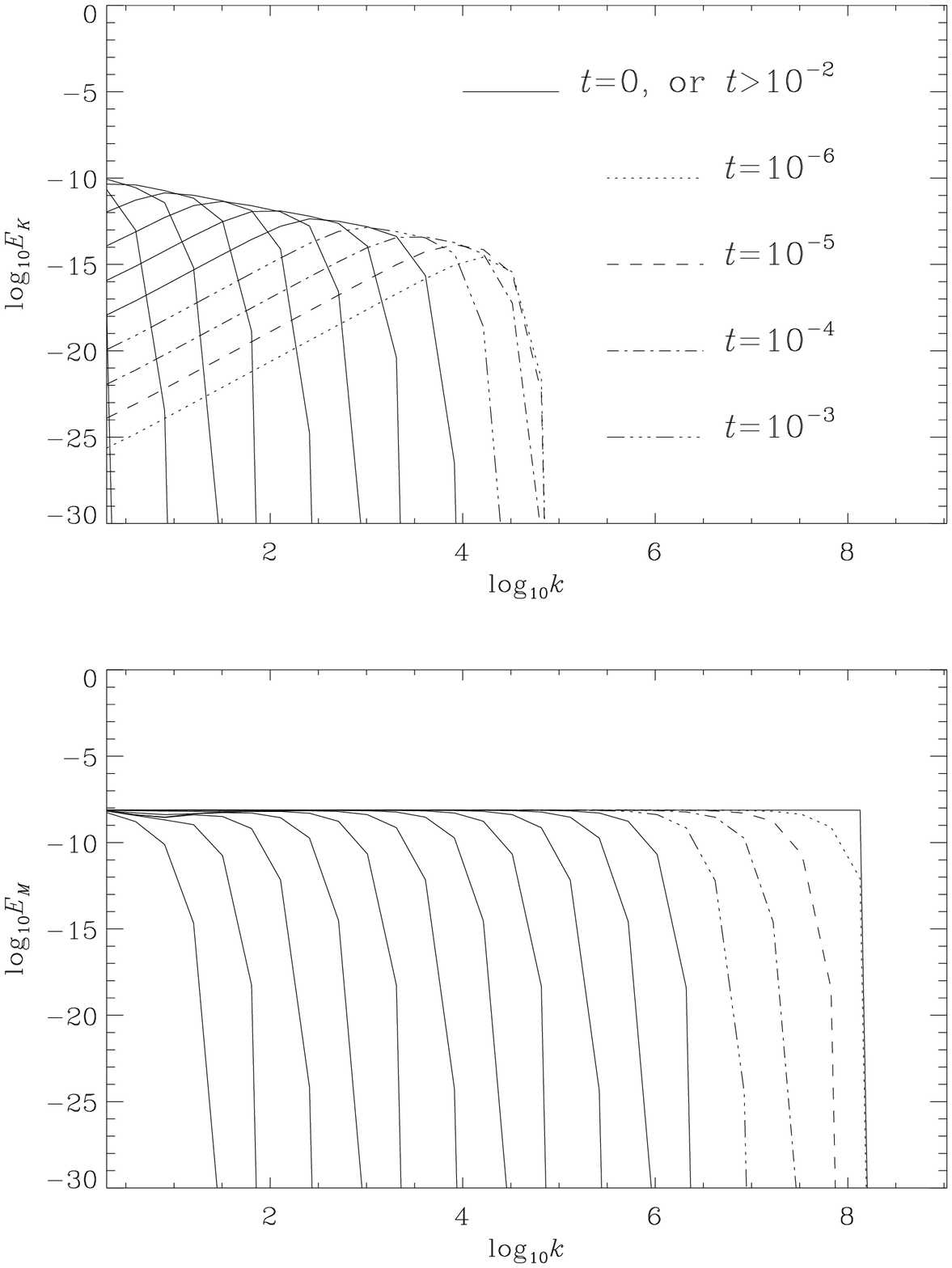}
\includegraphics{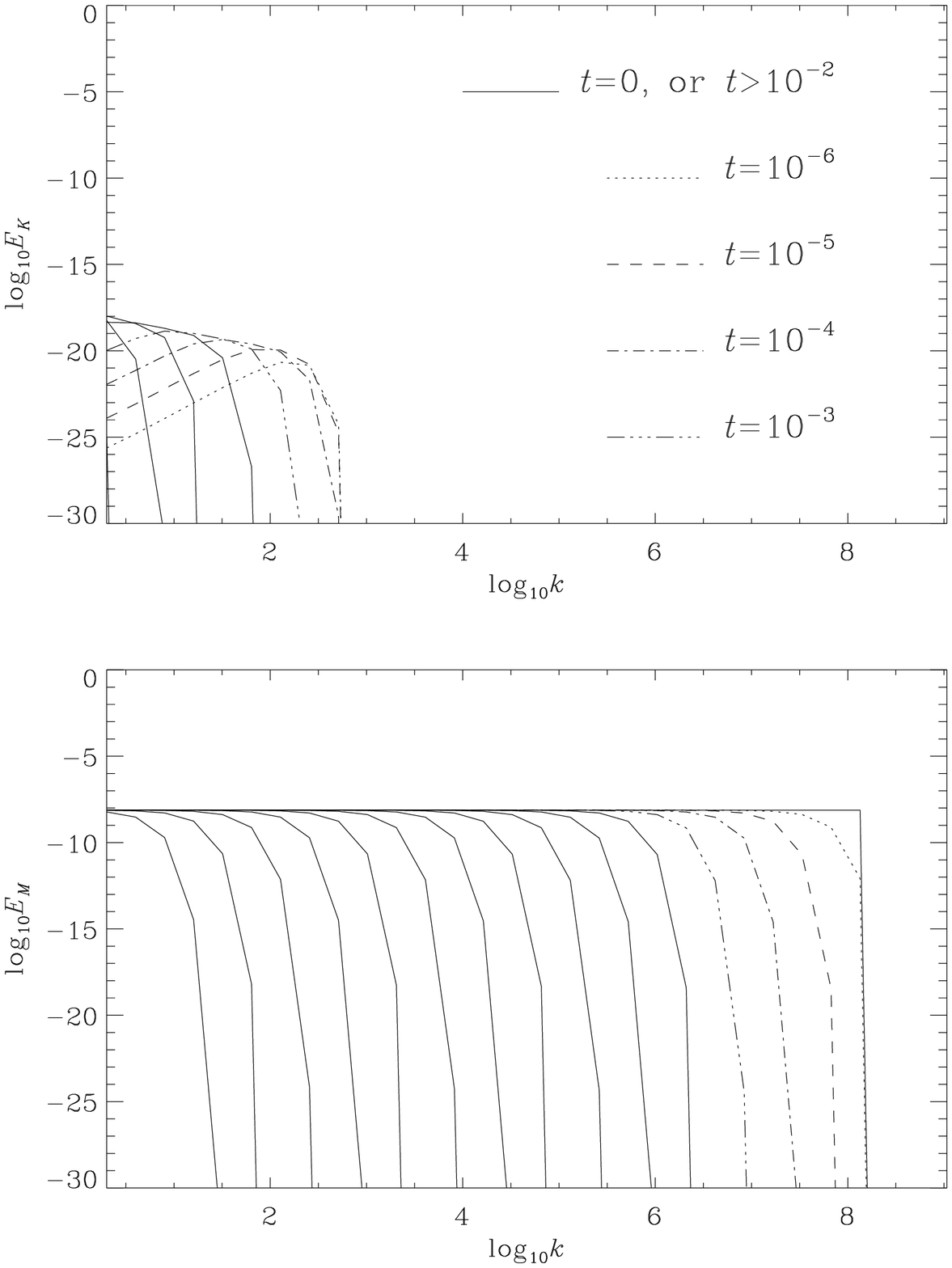}
\caption{Magnetic and plasma kinetic energy spectrum 
as a function of the wave number $k$ in the cascade model for  
small ($\nu=10^{-2}$) plasma viscosity (left) and large ($\nu=10^{2}$) 
plasma viscosity (right). The initial spectrum has been taken to be flat,
in accordance with the large-time behaviour of Fig. 1. 
The highest time $t=10^{8}$ corresponds to
the Hubble time $10^{16}.$}
\label{kuvaaxel}
\end{figure}

For a sufficiently large viscosity, the inverse cascade
stops. One may estimate \cite{beo2} 
that this typically takes place close to recombination.
The results suggest that in the real MHD, inverse cascade is
operative and is essentially not affected by Silk damping, except
very late and perhaps for very weak fields.
Thus we may conclude that it is unlikely that an equipartition
exists in the very early universe. A similar conclusion can be drawn in
a different, continuous  model where the inverse cascade can be found
analytically in an appropriate scaling regime \cite{beo2,poulin}.

\section{Conclusions}
Explaining the galactic magnetic fields in terms of microphysical
processes
that took place when the universe was only ten billionth of a second
old is a daunting task, which is not made easier by the complicated
evolution of the magnetic field as it is twisted and tangled
by the flow of plasma. It is nevertheless encouraging that 
mechanisms for generating primordial magnetic  fields of suitable size exist,
and in particular those based on the early cosmological
phase transitions discussed in Sect. 3 look promising.
At the same time the fact that there are so 
many possibilities tends to underline
our ignorance of the details of the subsequent evolution of the
magnetic field. The step from microphysics to macroscopic fields
is a difficult one because of the very large magnetic Reynolds number
of the early universe. However, different considerations, both analytic
approximations, 2d simulations, as well as the full-fledged shell
model computations which can account for turbulence, seem to point
to the existence of an inverse cascade of magnetic energy. Moreover,
as discussed in Sect. 4.3, the inverse cascade is obtained also in
the presence of a large plasma viscosity. Therefore the primordial
origin of the galactig magnetic fields is quite possible.

Much theoretical work remains to be done, though. At the same time
it is very important that progress is made on the observational
front. In particular, measuring or setting a stringent limit on the
intergalactic field, which could be possible in the near future
as indicated in Sect. 2.3, would provide the testing ground for  
all theoretical scenarios. 
\section*{Acknowledgments}
I wish to thank Poul Olesen for a fruitful
collaboration on primordial
magnetic fields, and Ola T\"ornkvist for useful discussions on bubble
collisions. This work has been supported by the Academy of Finland.
\newpage

\end{document}